\documentclass[aps,prl,twocolumn,superscriptaddress,showpacs]{revtex4}
\usepackage{amsmath,amsfonts,amssymb}
\usepackage{wrapfig}
\usepackage{graphicx}
\usepackage{epsfig}
\usepackage{bbm}
\usepackage{color}
\usepackage{multirow}

\newcommand{\ket}[1]{| #1 \rangle}
\newcommand{\bra}[1]{\langle #1 |}

\newcommand{\beq}{\begin{eqnarray}}
\newcommand{\eeq}{\end{eqnarray}}

\begin{document}

\title{Rerouting Excitation Transfer in the Fenna-Matthews-Olson Complex }
\author{Guang-Yin Chen\footnote{These authors contributed equally to this manuscript}}
\affiliation{Department of Physics and National Center for Theoretical Sciences, National
Cheng-Kung University, Tainan 701, Taiwan}
\affiliation{CEMS, RIKEN, Saitama, 351-0198, Japan}
\author{Neill Lambert$^*$}
\affiliation{CEMS, RIKEN, Saitama, 351-0198, Japan}
\author{Che-Ming Li}
\affiliation{Department of Engineering Science and Supercomputing Research Center, National Cheng-Kung University, Tainan
City 701, Taiwan}
\author{Yueh-Nan Chen}
\email{yuehnan@mail.ncku.edu.tw}
\affiliation{Department of Physics and National Center for Theoretical Sciences, National
Cheng-Kung University, Tainan 701, Taiwan}
\author{Franco Nori}
\affiliation{CEMS, RIKEN, Saitama, 351-0198, Japan}
\affiliation{Physics Department, University of
Michigan, Ann Arbor, MI 48109-1040, USA}
\date{\today }

\begin{abstract}
We investigate, using the Hierarchy method, the entanglement and the excitation transfer efficiency
of the Fenna-Matthews-Olson complex under two different local modifications: the suppression of transitions
between particular sites and localized changes to the protein environment. We
find that inhibiting the connection between the site-5 and site-6, or disconnecting site-5 from the complex completely, leads to
an dramatic enhancement of the entanglement between site-6 and site-7. Similarly, the transfer efficiency actually increases if site-5 is disconnected from the complex entirely. We further show that if site-5 and site-7 are conjointly removed, the efficiency falls. This suggests that while not contributing to the transport efficiency in a normal complex, site-5 introduces a redundant transport route in case of damage to site-7. Our results suggest an overall robustness of excitation energy transfer in the FMO complex under mutations, local defects,
and other abnormal situations.
\end{abstract}

\maketitle

Photosynthesis is one of the most important bio-chemical processes on earth %
\cite{van Amerongen}. When light is absorbed by a light-harvesting
antenna the excitation is transferred to a reaction center and used for
charge separation. Among the various photosynthetic complexes, the
Fenna-Matthew-Olson (FMO) complex in green sulfur bacteria is one of the
most widely studied \cite{Blankenship}. It has seven
electronically-coupled chromophores and functionally connects a large
light-harvesting antenna to the reaction center. Since the observation of
quantum coherent motion of an excitation within the FMO complex at 77 Kelvin %
\cite{Fleming}, considerable attention has been focused on the possible functional
role of quantum coherence in photosynthesis \cite{Lambert}. Recent
experiments further suggest the presence of quantum coherence even at room
temperature \cite{RT}.

Most quantum technologies, such as quantum computation, quantum
teleportation, quantum communication, rely on coherence in one way or another. Apart from
photonic qubits almost all physical realizations demand extremely
low-temperature environments to prevent fast dephasing \cite{Breuer} and
loss of quantum coherence. Therefore, the observation of quantum coherence
(entanglement) in the FMO complex at ambient temperature has naturally triggered
a great deal of theoretical interest and models \cite{Lambert, model, polaron1, airx2, aki, seth} focusing on
this biological system. The simplest theoretical treatment of the excitation
transfer in the FMO complex normally considers seven mutually coupled sites
(chromophores) and their interaction with the environment. One can either use
the Lindblad master equation, the more accurate Hierarchy method \cite{Hierarchy}, or
other open-quantum system models \cite{polaron1, polaron2, airx1, airx2} to
explain the presence of quantum coherence and predict the physical
quantities observed in experiments.

In a natural \emph{in-vivo} situation it is possible for the chromophores in the FMO complex to
suffer damage, e.g., from optical bleaching or mutation, such that a
transferring pathway is blocked, or such that the environment (protein) is modified in
some way. This has been demonstrated in recent experiments \cite{mutation}.
Motivated by this fact, we investigate in this work how the entanglement and
the transfer efficiency change when certain pathways are blocked, or the
properties of the local environment of one site are modified.  This question has been raised elsewhere, for example, Ref.~[\onlinecite{JMCP}] discusses, using a Markovian model, how various dissections of the FMO complex affect the efficiency and global entanglement.

Here, we specifically focus
on the situation where an excitation arrives at site-6, and must reach the reaction
center at site-3 (similar roles may be played by site-1 and site-4, respectively).  In this scenario, we ask the question
what role is played by site-5 (see Fig. 1), and what happens if it, or site-7, are damaged? We find that if site-5 is damaged or removed from the complex entirely,
the entanglement between sites 6 and 7 increases dramatically, as does the dynamic population of site-7 and consequently the
 efficiency (as characterized by the population of the `reaction centre')~\cite{olaya}.  We then show that if site-7 is damaged conjointly with site-5, the efficiency falls.  Thus, site-5, while not positively contributing
 to the efficiency in a perfect FMO complex, adds robustness and redundancy (as does the 6-1-2-3 transport route).

We begin with an brief introduction to the standard model of the FMO complex, and the description of the environment using the Hierarchy equations of motion.  We then discuss the concurrence and efficiency for damage and removal of site-5, and justify our interpretation of the role of site-5.  Finally, we also consider a simplified Markovian model of a 3-site system and obtain analytical results for the concurrence between two of the sites, to further elucidate our full numerical data.

\begin{figure}[th]
\includegraphics[width=\columnwidth]{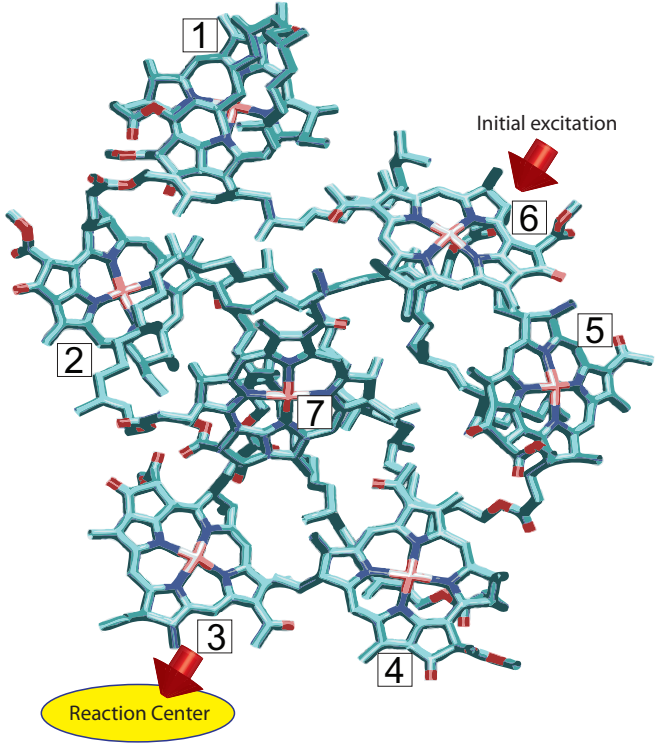}
\caption{(Color online) Schematic diagram of a monomer of the FMO
complex. The monomer consists of eight (only seven of them are presented
here) chromophores. The excitation (from the light-harvesting antenna)
arrives at sites 6 or 1 and then transfers from one chromophore to another.
When the excitation arrives at site 3, it can irreversibly move to the reaction
center. Here we assume that the initial excitation is at site 6.}
\end{figure}

\section{FMO Model}

Consider first a single FMO monomer containing $N=7$ sites, the general
Hamiltonian of which can be written as
\begin{equation}
H=\sum_{n=1}^{N}\epsilon _{n}|n\rangle \langle n|+\sum_{n<n^{\prime
}}J_{n,n^{\prime }}(|n\rangle \langle n^{\prime }|+|n^{\prime }\rangle
\langle n|)
\end{equation}%
where the state $|n\rangle $ represents an excitation at site $n$ ($n\in$ %
1,...,7), $\epsilon _{n}$ is the site energy of chromophore $n$, and $%
J_{n,n^{\prime }}$ is the excitonic coupling between the $n$-th and $n^{\prime }$%
-th sites.  Here, for simplicity, we omit the recently discovered eighth site \cite{olbrich} because its role on the excitation transfer process requires further studies.
It has been shown that the exitonic coupling $J_{n,n^{\prime }}$ is of the same
order as the reorganization energy, i.e., the coupling to the nuclear motion (phonons) of the protein environment. Thus a normal
secular Redfield, or Markovian Lindblad, treatment is insufficient~\cite{aki,NaturePhysics}, and the dynamics of the system must be modelled with a more complete
approach, such as the Hierarchy equations of motion~\cite{Hierarchy}.  These equations are non-perturbative and non-Markovian, and valid under the assumption of a Drude spectral density and an initially separable  system-bath state at $t=0$.  The Hierarchy is described by a set of coupled density matrices:
\beq
\dot{\rho}_{\mathbf{n}} &=& -\left(L + \sum_{j=1}^N\sum_{m=0}^K
\mathbf{n}_{j,m} \mu_m\right ) \rho_{\mathbf{n}} -
i\sum_{j=1}^N\sum_{m=0}^K\left[Q_j,\rho_{\mathbf{n}_{j,m}^+}\right]\nonumber\\
&-& i\sum_{j=1}^N\sum_{m=0}^K
n_{j,m}\left(c_mQ_j\rho_{\mathbf{n}_{j,m}^-} - c_m^*
\rho_{\mathbf{n}_{j,m}^-}Q_j\right). \eeq Here,
$Q_j=\ket{j}\bra{j}$ is the projector on the site $j$, $L$ is the
Liouvillian described by the Hamiltonian and the irreversible coupling to the reaction center (see below) $L=-\frac{i}{\hbar}[H,\rho_{\mathbf{n}}] +  L_{\text{sink}}$.  Here,
\beq C_j = \sum_{m=0}^{\infty} c_{j,m}
\exp\left(-\mu_{j,m} t\right) \eeq where $\mu_{j,0} = \gamma_j$,
$\mu_{j,m}\geq 1 = 2\pi m/\hbar \beta$, and the coefficients \beq
c_{j,0} = \gamma_j \lambda_j\left[\cot(\beta \hbar \gamma_j/2) -
i\right]/\hbar\eeq and \beq c_{j,m\geq 1} = \frac{4\lambda_j
\gamma_j}{\beta \hbar^2} \frac{\mu_{j,m}}{\mu_{j,m}^2
-\gamma_j^2}. \eeq
$\gamma_j$ is the
``Drude decay constant'', and indicates the memory time of the
bath for site $j$ (each site is assumed to have its own
independent bath), $\lambda_j$ is the reorganisation
energy, related to the system-bath coupling strength.

 A full description of the Hierarchy method can be found in the literature \cite{Hierarchy}, but in summary the Hierarchy is a large set of coupled equations each labelled by
$\mathbf{n}$, a set of non-negative integers uniquely specifying
each equation. The integers are defined as
$\mathbf{n}=\{n_1,n_2,n_3,...,n_N\} =
\{\{n_{10},n_{11},..,n_{1K}\},..,\{n_{N0},n_{N1},..,n_{NK}\}\}$.
In other words, each site $j$ has an additional label $m$, from $0$ to
$K$, and each of those labels in turn can run from $0$ to
$\infty$. The label $\mathbf{n}=0=\{\{0,0,0....\}\}$ is special,
and refers to the system density matrix.  Its properties at any
time $t$ define those of the system. This is in turn coupled to so-called ``auxiliary density matrices'', which describe the complex
bath fluctuations, by the terms in the equation with
$\mathbf{n}_{j,m}^{\pm}$ (i.e., $\mathbf{n}_{j,m}^{\pm}$ implies
the term in the index defined by $j$, and $m$ is increased or
decreased by $1$). At high temperature, and imposing the Ishizaki-Tanimura boundary condition \cite{Hierarchy}, we can cut the Hierarchy off at $K=0$ and an appropriate total number of terms in the remaining labels $N_c = \sum_{j,m} n_{j,m}$ providing convergence.

We also include $L_{\text{sink}}$ to describe the irreversible excitation transfer from
site 3 to the reaction center:
\begin{equation}
L_{\text{sink}}[\rho ]=\Gamma \lbrack \hat{s}\rho \hat{s}^{\dagger }-\frac{1}{2}\hat{s}^{\dagger }\hat{s}\rho
-\frac{1}{2}\rho \hat{s}^{\dagger }\hat{s}],
\end{equation}%
where $\hat{s}=|0\rangle \langle 3|$, with $|0\rangle $ denoting the state of the
reaction center, and $\Gamma $ the transfer rate.

In the FMO monomer, the excitation transfer from site 3 to the reaction center
occurs on a time scale of $\sim 1$ ps, and the dephasing occurs on a time
scale of $\sim 100$ fs \cite{seth}. This two time scales are both much
faster than that of the excitonic fluorescence relaxation ($\sim 1$ ns),
which, for simplicity, is omitted in our explicit results.

\begin{figure*}[th]
\includegraphics[width=2\columnwidth]{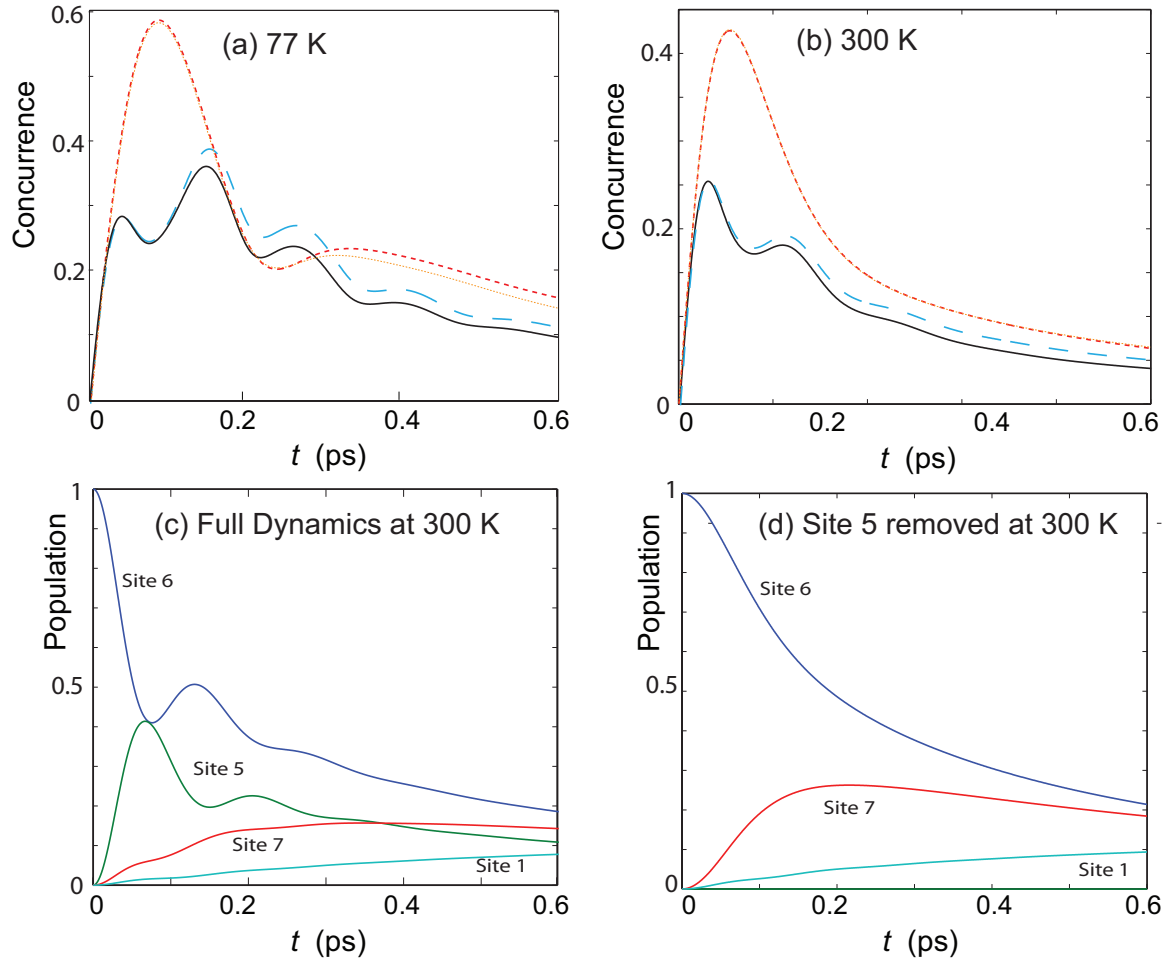}
\caption{(Color online) (a,b) The concurrence (coherence) of sites 6 and 7 for (a) a bath temperature of $77$ K and (b) $300$ K.
Not surprisingly, higher temperatures suppress oscillations. When the excitonic coupling between site-1 and site-6 (blue dashed curve) is inhibited the concurrence increases slightly. When the coupling between between site-5 and site-6 is inhibited (red dashed curve) or when site-5 is completely removed from the complex (orange dotted curve) the concurrence between 6 and 7 rises drastically. The solid black curve represents the
concurrence of the site-6 and site-7 for the full unmodified complex.
Figures (c,d) show the behavior of the populations at $300$ K.  Figure (c) shows the full unmodified dynamics, while (d) shows the case where site-5 is completely removed, and hence the population of $7$ rises at a faster rate. Interestingly, in (d) the coherent oscillations in the site-6 population disappear, while in (b) we see that the concurrence remains large, indicating that sometimes coherent oscillations are not a strong indicator of coherence (as also seen in \cite{NaturePhysics}).
In plotting this figure, we set $\gamma^{-1}~=~50$ fs and $\lambda~=~35$
cm$^{-1}$, and the rate from the site 3 to the reaction center $\Gamma^{-1}=1$ ps.}
\end{figure*}

\section{Concurrence and population dynamics in the presence of defects}

Each site in the FMO monomer may be decoupled from its nearest-neighbor sites
due to mutation-induced defects or rotation of the site \cite%
{mutation}. To investigate the effect of this change on the excitation
transfer we consider the situation where the initial excitation arrives at
site-6 and study the temporal excitation transfer when the excitonic
coupling between two specific sites is inhibited. We find that when the
coupling between site-5 and site-6 is inhibited, a significant enhancement
of the coherence between site-6 and site-7 can be obtained. To examine the
coherence between the two sites, we utilize the bipartite concurrence $C$,
which quantifies the degree of entanglement of any two sites $%
n $, $n^{\prime }$:\beq C_{n,n^{\prime }}=2|\bra{n}\rho _{\mathbf{0}}\ket{n'}|.\eeq  This is extracted from the $\rho_{\mathbf{0}}$ density matrix, evaluated from the time evolution of the Hierarchical equations of motion.

In Fig.~2(a,b), we show the concurrence $C$ of site-6 and site-7 when only the
excitonic coupling between site-1 and site-6 is inhibited [blue dashed line, for bath temperatures of $77$ K in (a) and $300$ K in (b)]. The
concurrence increases slightly since less sites share
the excitation from site-6. In contrast, when the coupling between the
just site-5 and site-6 is inhibited, a much larger enhancement (red dashed curve) of the
coherence can be obtained.  This is simply because the 5-6 coupling is much larger than the 6-1 coupling, thus when it is inhibited
more population can flow to site $7$, which, since it is a coherent process even at $300$ K, increases the concurrence between sites 6 and 7.  This can be further clarified with a simple three-site model, which we discuss in section IV \ref{3site}.

\subsection{Efficiency in the presence of defects}

What do these concurrence results imply for the overall efficiency of the transport process?  The efficiency of the excitation transfer can be formulated via the
population of the reaction center as a function of time,
\begin{equation}
P_{\text{RC}}(t)=\textrm{Tr}[\rho(t)\hat{s}\hat{s}^{\dagger}],
\end{equation}
where $\hat{s}=|0\rangle \langle3|$ is the operator connecting site-3 to the reaction center (see Fig.~1), denoted by the state $\ket{0}$, as defined earlier for the Lindblad $L_{\text{sink}}$.  Since the excitonic fluorescence relaxation of each individual site is slow  ($\sim 1$ ns) compared to all other time scales, $P_{\text{RC}}(t\rightarrow \infty)$ approaches unity, leading to the near $99$\% efficiency of the FMO complex commonly discussed in the literature.  It is often argued that coherence plays an important role promoting this high-efficiency, but some interesting investigations have shown that the quantum and classical models only differ by a few percent \cite{JianMa}.

To check this long-time behavior we employed an extended model (results not shown here), including the excitonic recombination rate of each individual site, and found that the defects discussed in the previous section do have a small effect on the long-time dynamics, and that the magnitude of this effect strongly depends on parameters which are not precisely known \cite{model}.  For example, a change in the rate $\Gamma$ between site-3 and the reaction center by a factor of $5$ results in a magnification of any differences in the efficiency. In addition, we found that any such change in the long-time dynamics is pre-determined by larger changes in the early-time population of the reaction center. Thus, here we use these short-time dynamics, in the absence of excitonic recombination, as an indicator of the efficiency.

In figure 3 we show the reaction center population as a function of time for a range of defects.  We see that, at both $77$ K and $300$ K, completely cutting site-5 (red dashed curve) enhances the reaction center population over the unmodified case (solid black curve), and hence enhances the efficiency. Conversely, cutting site-7 alone (long dashed blue curve) reduces the efficiency; in this case the population is forced to traverse through site-5, which is a less efficient, and slower, route to site $3$, and subsequently to the reaction center.  In contrast, removing both site-5 and site-7 (green dotted curve) leaves only the 6-1-2-3 transfer route, which due to the weak coupling between site 6 and 1 is less efficient.  This supports our earlier hypothesis that while site 5 does not contribute in a positive way to a perfect FMO complex, it does provide necessary redundancy in case of damage to the more efficient transport through site 7.

\begin{figure*}[th]
\includegraphics[width=2\columnwidth]{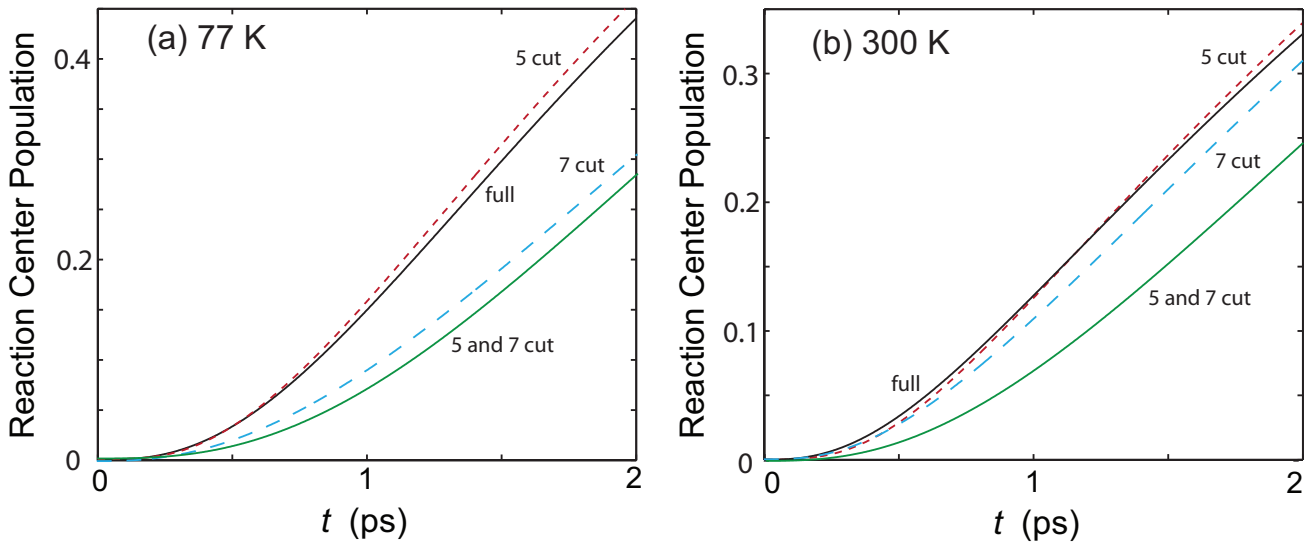}
\caption{(Color online) (a,b) The population of the reaction center (an indicator of the transport efficiency) for (a) a bath temperature of $77$ K and (b) $300$ K.
 When site-5 is removed from the complex (red dashed curve) the efficiency actually rises. When site-7 is removed (blue dashed curve), the population falls, while when both 5 and 7 are removed the population falls even further (green dotted curve).  The solid black curve represents the
population of the reaction center for the full unmodified complex.
As before, we set $\gamma^{-1}~=~50$ fs and $\lambda~=~35$
cm$^{-1}$, and the rate from the site 3 to the reaction center $\Gamma^{-1}=1$ ps.}
\end{figure*}

\section{Local Environmental Changes}

Other localized phenomena can also affect the transfer kinetics of
photosynthetic complexes. For example, the vibronic structure of the FMO
complex can be altered by the local substitution or deletion of the gene-encoding enzyme responsible for reducing the isoprenoid tail of the
chromophores \cite{mutation}. These alterations can lead to modifications
both of the protein and the chromophores.

We can easily investigate the
effect on the coherence when the local environment of one site is changed.
We assume that the modification of the local environment of site-5 results
in a stronger coupling to the protein environment. In Fig.~4, we show the concurrence $C$ of
site-6 and site-7 as the coupling site-5 to the environment, via the reorganisation energy $\lambda$, is increased.
In contrast to when site-5 was removed from the complex, the concurrence initially increases slightly, due to a small enhancement of the population flowing to site-7, and then decreases (with respect to the unmodified complex).  We also observed the change in the reaction center population as a function of these changes in $\lambda$ (not shown in the figure) and found an overall small increase, but not as drastic as that observed in Fig.~2.  The dynamics and efficiency, as a function of global changes in the environmental coupling and damping rate, have been well studied in many other works \cite{JianMa, aki}.  A full investigation, more rigorously taking into account the physical effect of the changes observed in \cite{mutation}, remains to be performed.

\begin{figure}[th]
\includegraphics[width=\columnwidth]{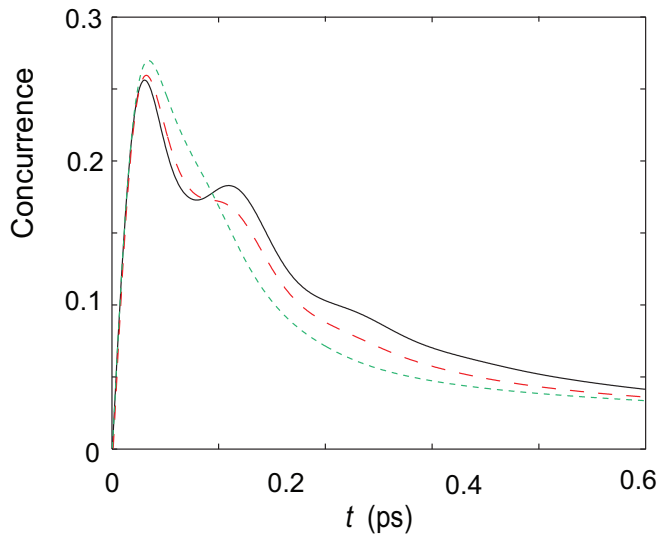}
\caption{(Color online) Concurrence of site-6 and site-7 for increased coupling of site-5 to the
protein environment, $\lambda_5=35$
cm$^{-1}$ (black solid), $\lambda_5=70$
cm$^{-1}$ (red dashed), and $\lambda=175$
cm$^{-1}$ (green dashed).  The bath temperature of all sites is kept at $300$ K, and $\gamma^{-1} = 50$ fs, $\Gamma^{-1}=1$ ps, as before. }
\end{figure}

\section{Three-site model}\label{3site}

To further understand the mechanism that leads to the enhancement of the
coherence between site-6 and site-7 observed in Fig.~2, we now consider a simplified three-site model, as shown in Fig.~5, where the three sites, 1, 2,
and 3, now represent site-6, -7, and -5, respectively, in the FMO complex. We assume that the initial
excitation is at site 1, and assign the inter-site couplings values
approximately corresponding to the excitonic couplings in the FMO monomer: $J_{1}=J_{6,7}$%
, $J_{2}=J_{5,7}$, and $J_{3}=J_{5,6}$. We further apply a Markovian
dissipative channel to both sites 2 and 3 with the rate $\gamma $ to
simulate the excitation transferring to other sites in the FMO monomer. Because $
J_{2}$ is small compared with $J_{1}$ and $J_{3}$, we can approximately set
$J_{2}=0$. For simplicity, we further assume $J_{1}=J$ and $J_{3}=\xi J$,
where $\xi $ is the tuning parameter for the coupling strength. The
concurrence of this simplified model can then be expressed as
\begin{eqnarray}
C &=&2\bigg|\frac{ie^{-t(\gamma +\Gamma )}(e^{\Gamma t}-1)J}{2\hbar\Gamma ^{3}}
\{\Gamma ^{2}+(\Gamma ^{2}-\gamma ^{2})e^{\Gamma t}  \notag \\
&+&\gamma \lbrack e^{\Gamma t}(\gamma +\Gamma )]\}\bigg|,
\end{eqnarray}%
where
\begin{equation}
\Gamma =\sqrt{-4(1+\xi ^{2})(J/\hbar)^{2}+\gamma ^{2}}.
\end{equation}

As shown in Fig.~5(a), the concurrence of sites 1 and 2 is strongly
enhanced (black solid curve) when the coupling $J_{3}$ between sites 1 and
3 is switched off. This is because the excitation population is
predominantly trapped between sites 1 and 2. In Fig. 5(b), we further show
how the concurrence increases while decreasing the coupling $J_{3}$.

\begin{figure}[th]
\includegraphics[width=7cm]{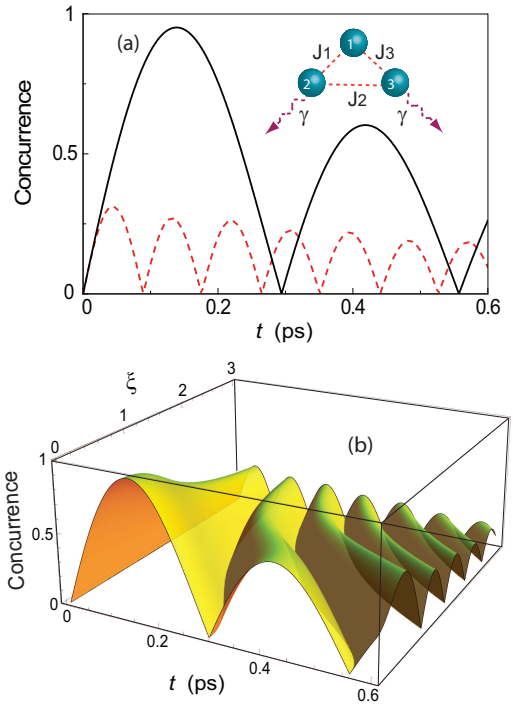}
\caption{(Color online) A simple model of three sites simulates the enhancement of the coherence
observed in the full model of the FMO complex in Fig.~2(a, b). Here the sites 1,2,
and 3 represent sites-6, -7, and -5 in the FMO monomer, respectively. The model
Hamiltonian has the same form (though now three sites are included) as Eq.
(1), and the inter-site couplings are set to be $J_{1}=J_{6,7}$, $%
J_{2}=J_{5,7}$, and $J_{3}=J_{5,6}$. Because $
J_{2}$ is small compared with $J_{1}$ and $J_{3}$, we approximately set
$J_{2}=0$. For simplicity, we further assume $J_{1}=J$ and $J_{3}=\xi J$
with $\xi $ being the tuning parameter. In order to simulate the excitation
transfer simply, we apply an additional Markovian dissipative channel with
rate $\protect\gamma $ to both sites 2 and 3. In (a) the red curve shows the concurrence between sites 1 and 2 for the full 3-site model.  The black curve shows
the concurrence when site-3 is decoupled from site-1. Figure (b) shows the
variation of the concurrence between site-1 and 2 as a function of time and
the parameter $\protect\xi $ of $J_{3}$. In plotting this figure, we apply $%
\protect\gamma~=~5.3~\text{cm}^{-1}$ and $J~=~30\text{cm}^{-1}$ .}
\end{figure}

\section{Conclusions}
In summary, we have shown that the concurrence (coherence) between site-6
and -7 can be enhanced significantly when the coupling between the sites-5
and -6 is inhibited, or if site-5 is removed from the complex completely.
In the latter case we also found a corresponding increase in the population of the reaction center.
We then argued that, rather than being superflous (and in fact contributing negatively to the efficiency of
a perfect FMO monomer), site-5 provides a backup in case of damage to the highly-efficient transport through site-7.
We further apply a simplified three-site model to
simulate this result.  Overall our results imply a robustness and redundancy to the energy
transfer in the FMO complex, as also noted in \cite{JMCP}.

This work is supported partially by the National Center for Theoretical
Sciences and National Science Council, Taiwan, grant number NSC
101-2628-M-006-003-MY3 and NSC 100-2112-M-006-017. F.N. acknowledges partial
support from the ARO, JSPS-RFBR Contract No. 12-02-92100, MEXT Kakenhi on
Quantum Cybernetics, and the JSPS-FIRST Program.

\end{document}